# Microwave gyroscope – novel rotation sensor

G.G.Karapetyan

Yerevan Physics Institute, Yerevan, Armenia

High performance microwave gyroscope (MG) is theoretically developed for the first time to our knowledge. MG is based on Sagnac effect in microwave ring resonator (RR), where a specially tailored phase shifter (PS) on the basis of surface acoustic waves is inserted. Due to that the beat frequency becomes proportional to square (or cubic) root upon rotation rate, and therefore hugely increases. In the result MG has few order higher sensitivity and dynamic range, than state-of-the-art laser gyros, so it can serve as an advanced rotation sensor in navigation, and fundamental sciences.

## 1. INTRODUCTION

Since the advent of laser three main types of optical rotation sensors have been under development. These include ring laser gyroscope (RLG), ring resonator gyroscope (RRG), and fiber optics gyroscope (FOG) [1-3]. State-of-the-art RLG is now the most sensitive device among them [4,5] and is used extensively in inertial navigation systems for aircraft, and in fundamental physics and geophysics [6]. All types of gyros are based on Sagnac effect, discovered in 1913 [7]. In Sagnac interferometer two light beams propagate in opposite directions around a common path. Rotating interferometer effectively shortens the optical path traveled by one of the beams, while lengthening the other on the same value $\delta L$, given by

$$\delta L = 2S\Omega / v . \qquad (1)$$

Here S -is the area enclosed by that path, v -is the speed of light there, $\Omega$ -is rotation angular frequency. In the case of FOG this change of path length causes phase difference between two counter propagating beams (having the same frequency), which is detected by interfering them outside the path. However in RLG and RRG light propagates in RR so change of optical path length causes the change of RR resonant frequency. Thus, counter propagating beams in RR have different frequencies, and produce the beat frequency $2\delta f$, which is direct proportional to change of path length

$$2\delta f = -2f\delta L / L = -4fS\Omega / vL , \qquad (2)$$

where f – is frequency of light.

As it is seen from (2) beat frequency is direct proportional also to operating frequency, so a microwave gyroscope with operating frequency around 1 GHz will have about on 5 orders lower beat frequency, and therefore sensitivity than RLG with the same dimensions. Because of that till now microwave gyros have not been considering as a possible rotation sensors at all. Now we use a novel method, called *the method of phase shifting* enabling to increase considerably the beat frequency in rotating RR. Due to that a possibility of creating an advanced rotation sensor with very high performance - microwave gyroscope arises.

## 2. PRINCIPAL EXPRESSIONS

Let us consider a microwave RR, with perimeter L. Its resonant frequencies $f_0$ are determined from the condition that path length is multiple to the wavelength, or that along that path the phase of microwaves is changed on an integer $2\pi$, i.e.

$$Lf_0 / v = m , \qquad (3)$$

where m – is an integer.

Suppose that a frequency dependent PS, which adds the phase $\varphi(f)$ to the waves with frequency f is inserted into RR. Then the resonant frequencies $f_1$ of that RR satisfy to another condition

$$Lf_1 / v + \varphi(f_1) / 2\pi = m . \qquad (4)$$

When RR is rotated path length of waves are changed on a small value $\Delta L$, which causes the shift $\Delta f$ of resonant frequency, determined from the equation

$$(L+\Delta L)(f_1 + \Delta f)/v + \varphi(f_1 + \Delta f)/2\pi = m. \qquad (5)$$

Let us assume that φ(f) is a quadratic function in vicinity of point f=$f_1$. Then φ($f_1$+Δf) can be expressed as an expansion in Taylor series with 3 terms:

$$\varphi(f_1 + \Delta f) = \varphi(f_1) + \varphi' \Delta f + \varphi'' \Delta f^2 /2, \qquad (6)$$

where primes mean derivatives in respect to the frequency with arguments $f_1$.
Substituting (6) in (5) we have

$$\varphi'' \Delta f^2 + 2[2\pi(L+\Delta L)/v + \varphi']\Delta f + 4\pi f_1 \Delta L/v = 0. \qquad (7)$$

This square equation determines the shift of RR resonant frequency stipulated by the shift of its length. The solutions of (7) is written as

$$\Delta f = \left(-A \pm \left(A^2 - 4\pi f_1 \varphi'' \Delta L/v\right)^{1/2}\right)/\varphi'', \qquad (8)$$

where A=2π(L+ΔL)/v+φ′.
Analyzing (8) one can conclude that shift of RR resonant frequency is sharply increased when φ′→ −2πL/v, becoming proportional to square root of ΔL (see Fig.1, and Fig.2). By this it is necessary that ΔL and φ″ have opposite signs, because if φ″ΔL>0 then (8) gives imaginary value of Δf, which indicates on the absence of oscillations in RR. Therefore in rotating RR with PS one of counter propagating waves is disappeared, but another one is splitted on two waves propagating in the same directions. Below we will assume that condition φ″ΔL< 0 is satisfied. Thus appropriately installed negative value of PFD can strongly change the functional dependence between Δf and ΔL, considerably increasing Δf. The range of PFD, where it takes place is determined from (8) by taking into account the condition $A^2$<<$f_1$φ″ΔL /v, which leads to

$$|\varphi' + 2\pi L/v| < (f_1 |\varphi'' \Delta L|/v)^{1/2}. \qquad (9)$$

By satisfying this condition the shifts of RR resonant frequency are

$$\Delta f \approx \pm(-4\pi f_1 \Delta L/v\varphi'')^{1/2} = \pm f_1 (8\pi S\Omega/|\varphi''|)^{1/2}/v, \qquad (10)$$

that produce the beat frequency 2Δf, considerably surpassing the beat frequency (2) in RR without PS, because of its proportionality to square root upon rotation rate. By analogy a PS with cubic function φ(f) in vicinity of point f=$f_1$ can be considered. Then φ″=0, and Δf is determined by third derivative φ‴

$$\Delta f \approx -(12\pi f_1 \Delta L/v\varphi''')^{1/3} \approx -f_1(-24\pi f_1 S\Omega/v^2 \varphi''')^{1/3}. \qquad (11)$$

Contrary to previous case, both of counter-propagating waves exist here. One of them has positive shift of resonant frequency, another one- negative. Beat frequency produced by them is however higher than that in previous case because of its proportionality to cubic root upon rotation rate.
Required PS can be designed on the basis of surface acoustic waves (SAW) delay lines [8] (with bi-directional amplifier to compensate the losses). Since SAW wavelength λ is around a micron the change s in path traveled by SAW for example on 0.1 mm causes the phase shift ~s/λ~100. To obtain required by (9) value of PFD it is necessary that this phase shift took place in frequency interval Δf~(v/L)(s/λ). Thus required PFD can be obtained by tailoring appropriate chirp low of SAW transducers. Following this approach we designed a PS on the basis of SAW chirped delay line. Its transducers have both linear and quadratic chirp, providing the quadratic function φ(f) in vicinity of resonant frequency 600 MHz, having φ′= −0.09424 1/MHz, and φ″=0.0001 1/MHz$^2$.

### 3. DISCUSSION
Let us compare the performances of proposed MG and optical gyros. Numerical evaluations of (10) by substituting L=3 m, v=2·$10^8$ m/s, φ″=0.0001 1/MHz$^2$, $f_1$=600 MHz is presented on Fig. 3 by solid line. The beat frequency in conventional MG without PS, determined from (2) is plotted by dashed line. By dotted line it is presented the beat frequency in RLG having the same dimensions. It is seen that after inserting of PS, the beat frequency, and therefore sensitivity of MG increases on 7…9 order, and surpasses by this a few order the beat frequency in RLG. Dynamic range of MG also is much larger than that of RLG, because change of rotation rate in region for example $10^{10}$ causes the change of beat frequency only in region $10^5$. Therefore proposed MG is an advanced rotation sensor, that being created can replace existing RLG and FOG in inertial navigation systems because of its higher performance, and possible lower cost. Another application of MG can be monitoring of Earth rotation. As it follows from (10) Earth rotation angular velocity 15 grad/hour causes the beat frequency of MG (Earth Sagnac frequency) about 800 Hz, meanwhile the world largest laser gyros – Canterbury C-2 produces Earth Sagnac frequency only 79 Hz [9]. Moreover, beat frequency 800 Hz can be increased further by increasing of MG dimensions, and operating frequency. Thus MG can be used also in geophysics for monitoring Earth rotation angular velocity with very high precision.

## 4. CONCLUSIONS

In conclusion we proposed an advanced rotation sensor - MG, where appropriately tailored SAW chirped delay line is inserted. Due to that beat frequency becomes proportional to square (or cubic) root upon rotation rate so is hugely increased, surpassing even the beat frequency in RLG with the same dimensions. In the result MG has higher sensitivity and dynamic range than state-of-the-art RLG. Being created proposed MG can be used in inertial navigation systems, and fundamental sciences.

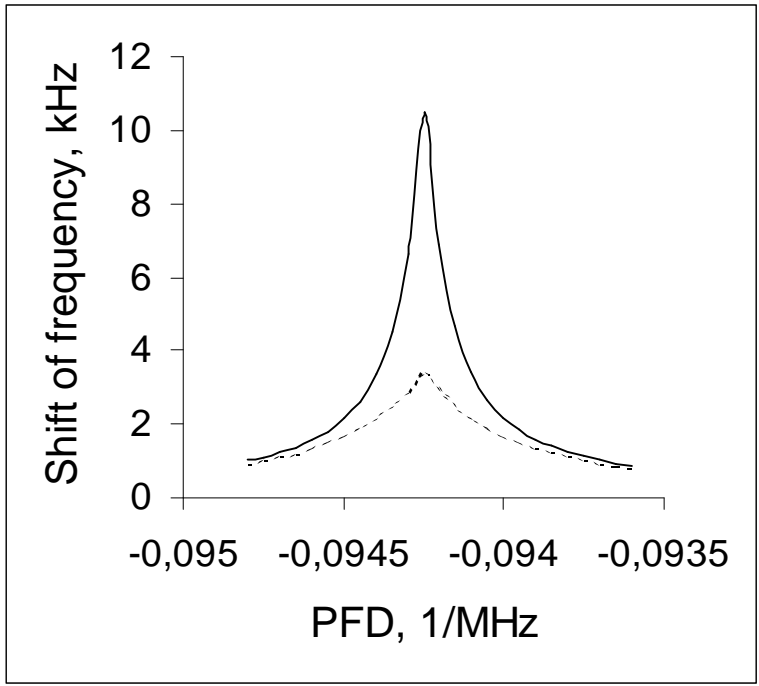

Fig.1 Frequency shift in RR versus PFD of phase shifter, when $\Delta L$=30 nm, L=3 m. The solid line corresponds to $\varphi''$=0.01 1/MHz$^2$, the dotted line $\varphi''$=0.1 1/MHz$^2$.

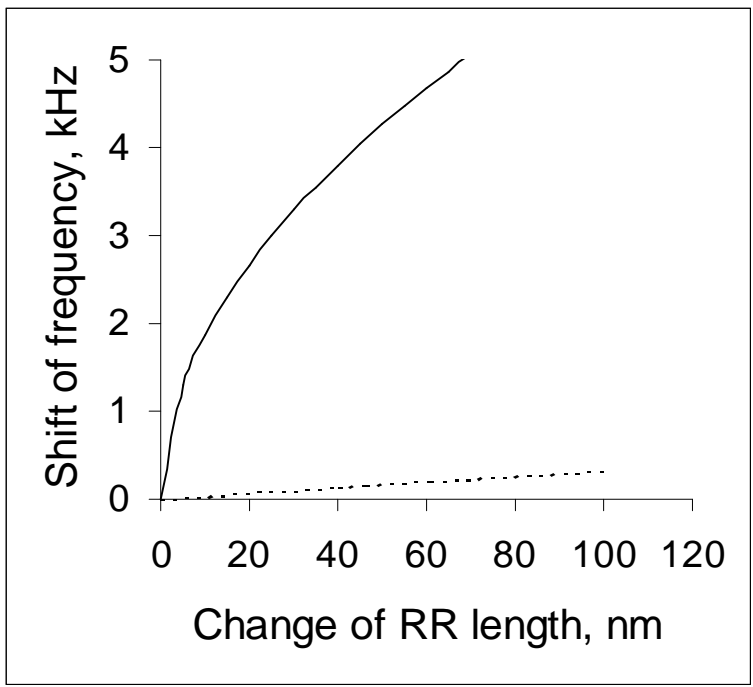

Fig.2 Frequency shift versus the shift of RR length, when $\varphi'' = -0.1$ 1/MHz$^2$, L=3 m. The solid line corresponds to $\varphi' = -0.09424$ 1/MHz, the dotted line $\varphi' = -0.1$ 1/MHz.

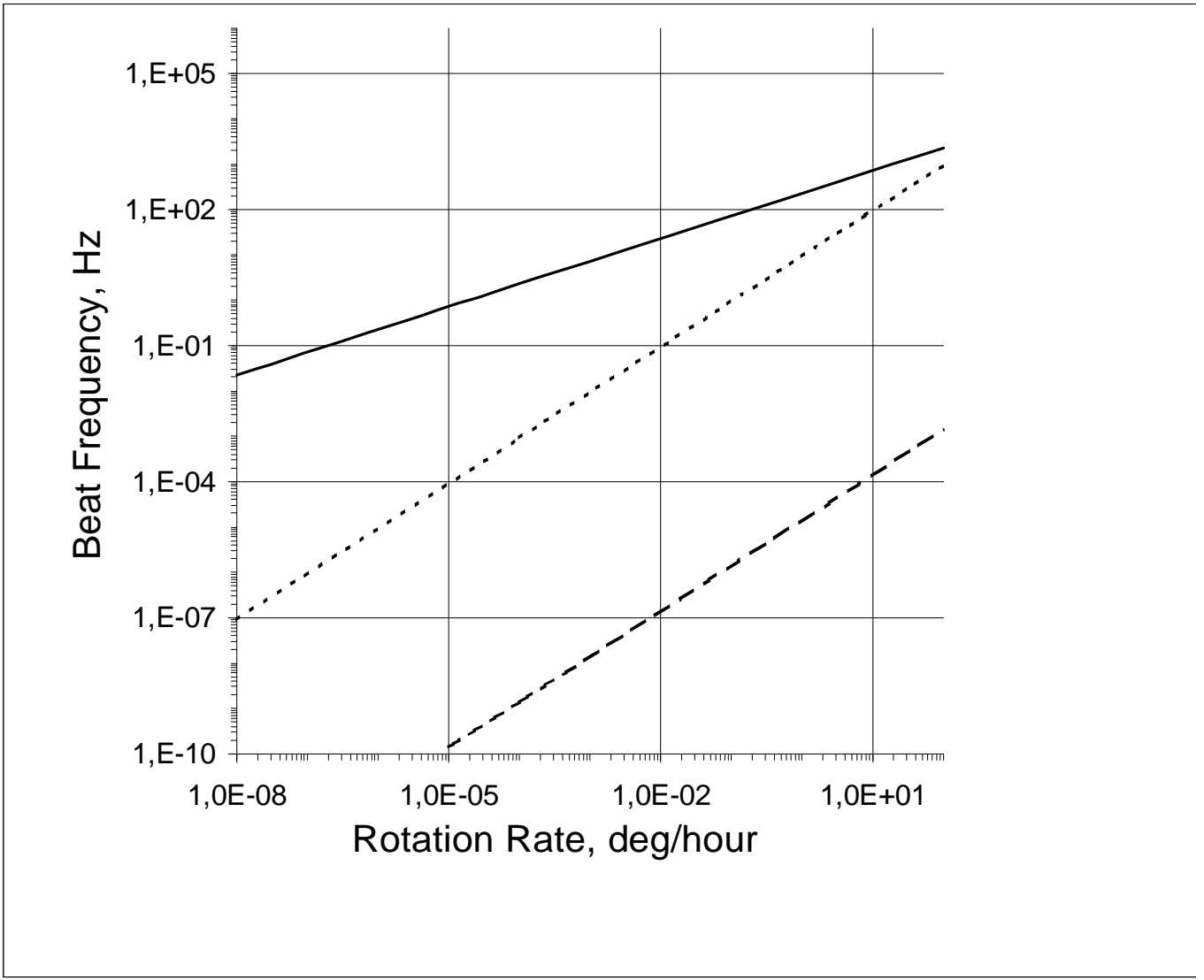

Fig.3 Beat frequency in conventional MG (the dashed line), and in MG with PS (the solid line). L=3m, v=2x10$^8$m/s, f$_1$=600 MHz, φ″=0.0001 1/MHz$^2$. The dotted line corresponds to RLG having L=3m, v=3x10$^8$m/s, f$_1$=6x10$^{14}$ Hz.